\documentclass[runningheads]{llncs}
\usepackage{graphicx}
\usepackage{multirow}
\usepackage{url}
\begin{document}
\title{Survival prediction using ensemble tumor segmentation and transfer learning}
%
%
\author{Mariano Cabezas\inst{1} \and
Sergi Valverde\inst{1} \and
Sandra Gonz\'alez-Vill\`a\inst{1} \and
Albert Cl\'erigues\inst{1} \and
Mostafa Salem\inst{1} \and
Kaisar Kushibar\inst{1} \and
Jose Bernal\inst{1} \and
Arnau Oliver\inst{1} \and
Xavier Llad\'o\inst{1}
}
\authorrunning{Mariano Cabezas et al.}
%
\institute{Research Institute of Computer Vision and Robotics, University of Girona, Spain\\\email{mcabezas@eia.udg.edu}}
\maketitle              
\begin{abstract}
Segmenting tumors and their subregions is a challenging task as demonstrated by the annual BraTS challenge. Moreover, predicting the survival of the patient using mainly imaging features, while being a desirable outcome to evaluate the treatment of the patient, it is also a difficult task. In this paper, we present a cascaded pipeline to segment the tumor and its subregions and then we use these results and other clinical features together with image features coming from a pretrained VGG-16 network to predict the survival of the patient. Preliminary results with the training and validation dataset show a promising start in terms of segmentation, while the prediction values could be improved with further testing on the feature extraction part of the network.

\end{abstract}
\section{Introduction}
Gliomas are the most common primary brain malignancies, with different degrees of aggressiveness, variable prognosis and various heterogeneous histological sub-regions, i.e. peritumoral edema, necrotic core, enhancing and non-enhancing tumor core. This intrinsic heterogeneity of gliomas is also portrayed in their imaging phenotype (appearance and shape), as their sub-regions are described by varying intensity profiles disseminated across multimodal MRI scans, reflecting varying tumor biological properties. Due to this highly heterogeneous appearance and shape, segmentation of brain tumors in multimodal MRI scans is one of the most challenging tasks in medical image analysis.

The BraTS challenge~\cite{Menze2015,Bakas2017,BraTSa,BraTSb} started on 2013 and since then, it has always been focusing on the evaluation of state-of-the-art methods for the segmentation of brain tumors in multimodal magnetic resonance imaging (MRI) scans. The current iteration, BraTS 2018 utilizes multi-institutional pre-operative MRI scans and focuses on the segmentation of intrinsically heterogeneous (in appearance, shape, and histology) brain tumors, namely gliomas. Furthemore, to pinpoint the clinical relevance of this segmentation task, BraTS’18 also focuses on the prediction of patient overall survival using radiomic features and automatic machine learning algorithms.

On BraTS 2017, Kamnitsas et al.~\cite{Kamnitsas2017b} obtained the best segmentation results with significant differences with respect to the other challengers. This approach introduced the use of an ensemble classifier composed of different deep convolutional neural networks (CNN) architectures. Moreover, in terms of survival prediction the best approach by Shboul et al. using a random forest approach after a previous feature extraction step where image features are extracted from a segmented image. In this paper we present a new approach based on these two approaches where we first presegment the tumor region using a 3D unet~\cite{Ronneberger2015} and then we refine this segmentation with a cascade approach using an ensemble of 4 different CNN architectures. Finally, using the VGG-16~\cite{Simonyan2014} network pretrained on the imagenet dataset, we extract features on 20 tumor slices and combine them with the clinical data to obtain a survival prediction. The rest of the paper is structured as follows: in section~\ref{sec:methods} we present our approach for each task, followed by the preliminary results on the training and validation dataset and a discussion in section~\ref{sec:results}. Finally, our conclusions are presented in~\ref{sec:conclusions}.

\section{Methods}
\label{sec:methods}
\subsection{Task 1: Segmentation of gliomas in pre-operative MRI scans}
Following our approach for multiple sclerosis segmentation by Valverder et al.~\cite{Valverde2017} and last's year's submission on the BraTS challenge by Wang et al.~cite{Wang2017} we decided to implement a cascaded approach. However, unlike these two approaches, we used two different networks for each step. 

\subsubsection{Tumor delineation}
\begin{figure}[t]
    \begin{center}
    \includegraphics[width=\textwidth]{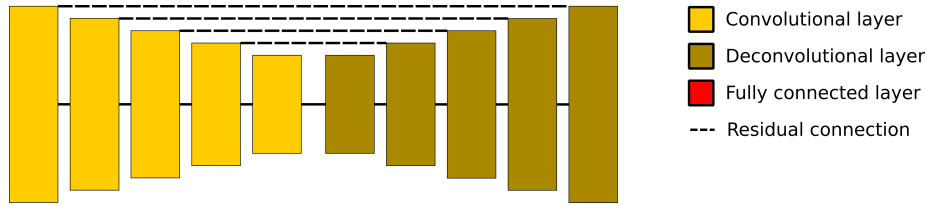}
    \end{center}
    \caption[3D unet architecture for segmentation]{3D unet architecture for segmentation. This network uses only convolutional and deconvolutional layers of 32 filters with a kernel size of 3 $\times$ 3 $\times$ 3. It also includes residual connections between convolutional and deconvolutional layers.}
    \label{fig:net1}
\end{figure}

Fully convolutional networks, and specifically unet ones, have a high accuracy when segmenting lesions with a small amount of data due to the capibilities of using large blocks of input data to train each convolutional kernel. However, they usually have poor results when trying to segment small subregions from a given mask. Therefore, we used a unet network to first delineate the tumor ROI mask.

This network (see figure~\ref{fig:net1} uses 5 levels of convolutional and deconvolutional layers of 32 filters with a kernel size of 3 $\times$ 3 $\times$ 3. Residual connections were also used to improve gradient decay, and the final output is given by a convolutional layer with kernel size of 1 $\times$ 1 $\times$ 1 followed by a softmax activation. As input, patches of size 21 $\times$ 21 $\times$ 21 were used for training, while the whole image was used for testing, since the network is fully convolutional and it speeds up the process.

\subsubsection{Subregion segmentation}
\begin{figure}[t]
    \begin{center}
    \includegraphics[width=\textwidth]{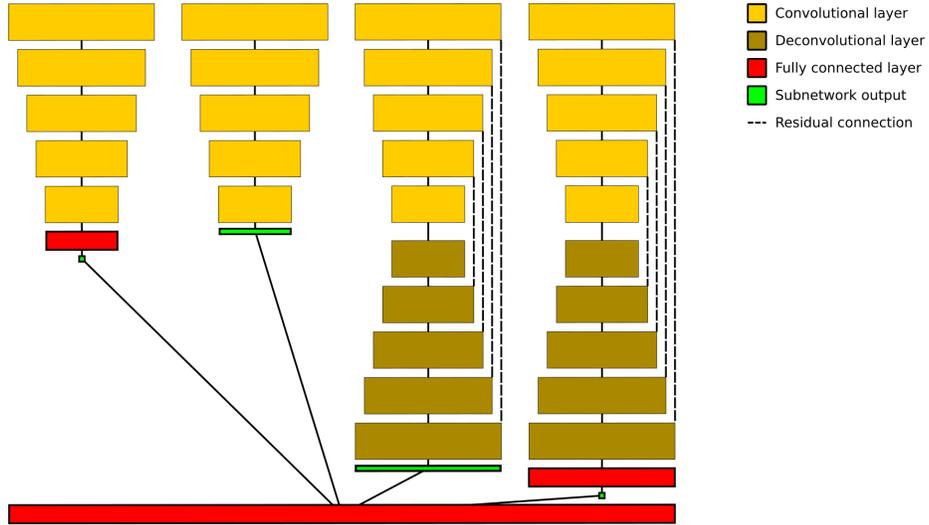}
    \end{center}
    \caption[Ensemble network for segmentation]{Ensemble network for segmentation. The different networks use the same kind of convolutional and deconvolutional layers (represented in light yellow and dark yellow respectively) with 32 filters and kernels of 3 $\times$ 3 $\times$ 3 and the fully connected layers all have 5 4 units (for the 3 tumor subregions + the background labels).}
    \label{fig:net2}
\end{figure}
With this initial ROI segmentation, we then apply a second network focused on segmenting the tumor subregions. Following Kamnitsas et al.~\cite{Kamnitsas2017b}, we implement a small ensemble net. However, instead of using previous architectures, we propose a framework where each small subnet shares some metaparameters, while having different overall architectures and weights. The goal is to capture each network's bias but keeping the same input information. The architectures include a unet with a dense output (UCNN), a unet with a fully convolutional output (UNET) and a dense (CNN) and fully convolutional networks (FCNN) using only convolutions (as illustrated by figure~\ref{fig:net2}.

These networks were trained independently using the same input patches of size 13 $\times$ 13 $\times$ 13 to guarantee that the final convolutional later of both the CNN and FCNN networks were of size 3 $\times$ 3 $\times$ 3 (to reduce the number of parameters of the final dense layer of the CNN network). Finally, a dense layer was trained with the output of the previous networks to give the ensemble results instead of using the average of all the networks. This training was performed on two steps, first for the indepent networks (to avoid expanding their biases into each other) and then for the last dense layer with the previous networks frozen.

\subsection{Task 2: Prediction of patient overall survival (OS) from pre-operative scans}
After segmenting the tumor subsections, we use their volumes as features together with the clinical data (age) as features for the survival task. Moreover, we decided to also include image features based on the surrounding area of the tumor.

\begin{figure}[t]
    \begin{center}
    \includegraphics[width=\textwidth]{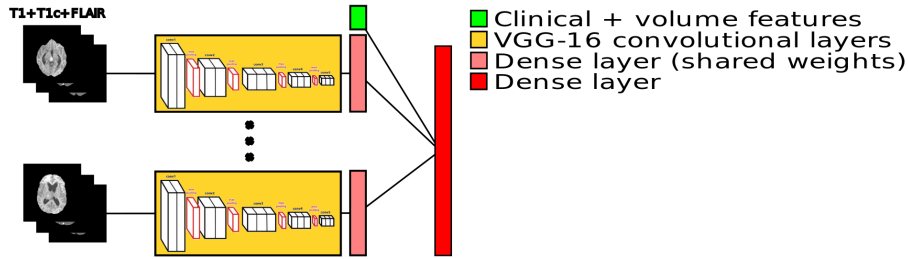}
    \end{center}
    \caption[Survival prediction architecture]{Survival architecture composed by a pretrained VGG-16 network, a dense layer with weight shared for each slice and a final layer that takes also into account clinical and volumetric features.}
    \label{fig:survival_net}
\end{figure}

Due to promising results on transfer learning tasks using pretrained networks on natural images, we decided to use the VGG-16 network pretrained with the ImageNet dataset to compute features for 20 slices around the center of the tumor. These features are then passed through a fully convolutional layer (that shares weights among all the slices) of 156 units. Finally, the output of this layer is then combined with the clinical and volume features to obtain a final survival prediction as illustrated by figure~\ref{fig:survival_net}.

\subsection{Implementation details}
All the work was developed using the Keras library and python 2.7. Moreover, it was tested on a NVIDIA GTX Titan Xp and a Titan X PASCAL gpus with 12GB of RAM and a total system RAM of 256GB. Finally, all the code is publicly available at:~\url{https://github.com/marianocabezas/challenges2018}.

\section{Results and discussion}
\label{sec:results}
\subsection{Task 1: Segmentation of gliomas in pre-operative MRI scans}
In order to only evaluate the final results and avoid sending multiple submissions to the website, we only present the quantitative results of segmentation for the final ensemble in table~\ref{tab:segmentation}.

\begin{table}[ht]
\scriptsize
\centering
\caption{Summary of the results obtained from the IPP evaluation website for the segmentation task.}
\label{tab:segmentation}
\begin{tabular}{|c||cc||cc|}
\hline
\multirow{2}{*}{Metric} & \multicolumn{2}{c||}{Training} & \multicolumn{2}{c|}{Validation} \\
                        & Mean ($\pm \sigma$) &  Median  &  Mean ($\pm \sigma$) & Median   \\
\hline
\hline
Dice ET                 & 0.66721 (0.29115)   & 0.78909  & 0.74034 (0.27735)    & 0.8339   \\
Dice WT                 & 0.84913 (0.13001)   & 0.89387  & 0.88928 (0.07497)    & 0.91197  \\
Dice TC                 & 0.71729 (0.23442)   & 0.80191  & 0.72644 (0.24267)    & 0.80026  \\
Sensitivity ET          & 0.7343 (0.25561)    & 0.80397  & 0.76934 (0.26688)    & 0.86809  \\
Sensitivity WT          & 0.80368 (0.17829)   & 0.8604   & 0.88843 (0.11746)    & 0.94084  \\
Sensitivity TC          & 0.70196 (0.26795)   & 0.79212  & 0.76289 (0.2573)     & 0.87682  \\
Specificity ET          & 0.99762 (0.00432)   & 0.99877  & 0.99803 (0.00285)    & 0.99886  \\
Specificity WT          & 0.99649 (0.00416)   & 0.99793  & 0.99483 (0.00381)    & 0.99606  \\
Specificity TC          & 0.99613 (0.00677)   & 0.99844  & 0.99529 (0.00607)    & 0.99735  \\
Hausdorff95 ET          & 7.59116 (10.95449)  & 2.65791  & 5.3035 (9.96395)     & 2.23607  \\
Hausdorff95 WT          & 7.76567 (10.008)    & 4.89898  & 6.95631 (11.9391)    & 3.31662  \\
Hausdorff95 TC          & 9.82279 (8.50929)   & 8.06226  & 11.92386 (13.44799)  & 8        \\
\hline
\end{tabular}
\end{table}

Looking at the table, the results seem to be consistent between both datasets, even though the validation set provides slightly better results. This might be related to the fact that the number of cases (and most likely their variability) is lower in the validation dataset due to a lower number of cases.

Looking at the differences between the mean results and the median ones, we observer how the median is higher for both datasets. That also suggests, that while the overall performance is good, there are some outlier cases where the performance is worsened. For instance, this is the case for patient Brats18\_CBICA\_AOH\_1 from the training dataset. For this patient, the DSC values for enhancing tumor (ET), whole tumor (WT) and tumor core (TC) are 0, 0.43607 and 0 respectively. In fact, for some cases of the training and validation dataseta DSC value of 0 was obtained for the enhancing tumor region, proving to be one of the most difficult.

\begin{figure}[p]
    \begin{center}
    \includegraphics[width=\textwidth]{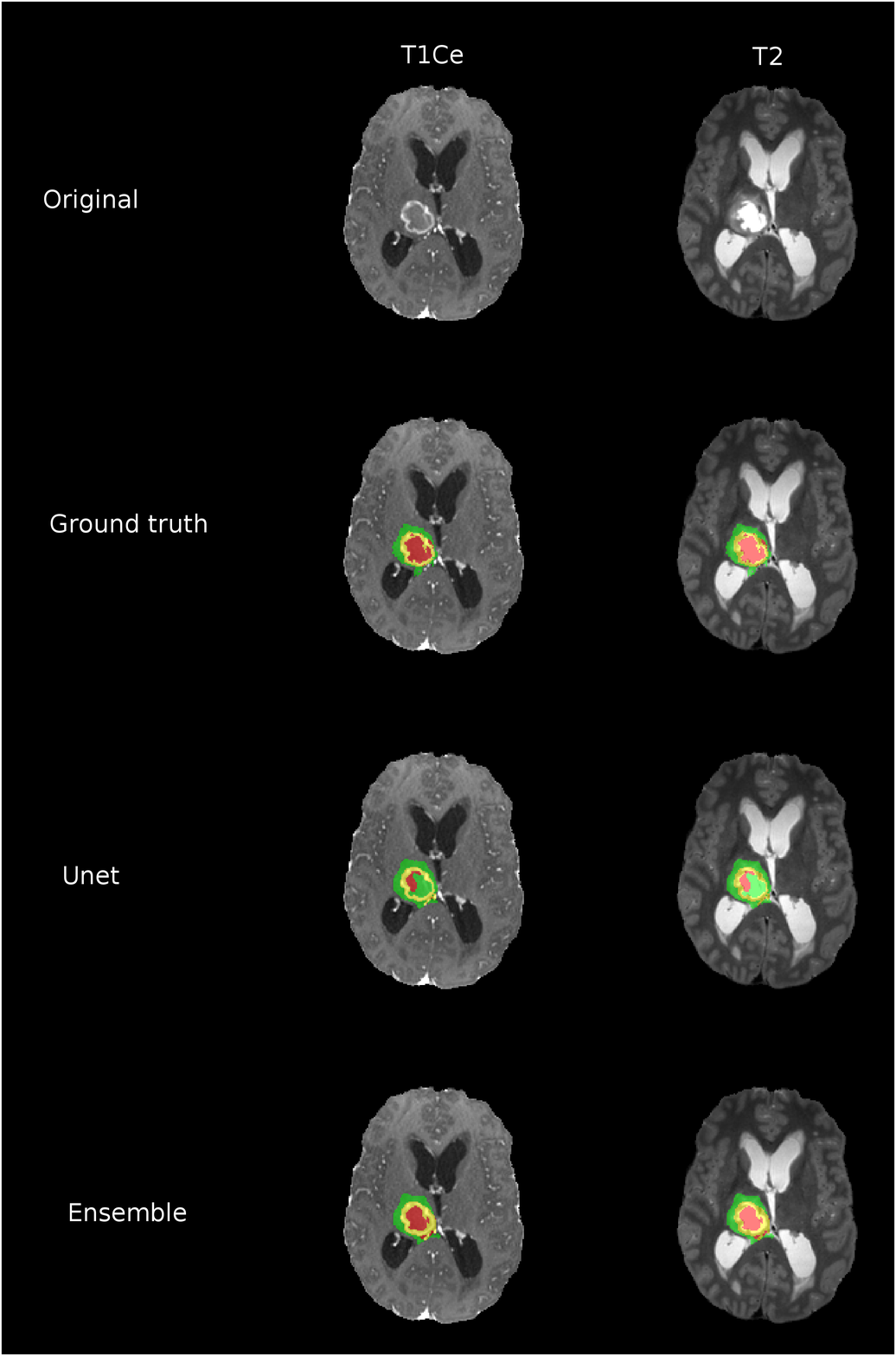}
    \end{center}
    \caption[Qualitative example]{Qualitative example of the segmentation of one slice. While the unet provides a good ROI segmentation, the subregions are not well delineated, while the ensemble improves this segmentation.}
    \label{fig:qualitative}
\end{figure}

In order to show the improvement of the cascaded approach, some qualitative examples are presented in figure~\ref{fig:qualitative}. As observed in this figure, the overall ROI of the tumor is similar between both networks. However, while the unet failed to properly differentiate between tumor subregions, the ensemble (which has two subnetwork dedicated to classify the central voxel of a patch) is capable of better delineate these subregions, as expected. Moreover, we can clearly see how the unet undersegmented the enhancing tumor area. This strengthens the notion that this is the most challenging tumor region to segment.

\subsection{Task 2: Prediction of patient overall survival (OS) from pre-operative scans}
The results obtained for the survival task are presented on table~\ref{tab:survival}. As observed in the table, the results are fairly low. In fact the accuracy is below 0.5 for both validation and training. However, there are a few issues to take into account.

\begin{table}[ht]
\scriptsize
\centering
\caption{Summary of the results obtained from the IPP evaluation website for the survival task.}
\label{tab:survival}
\begin{tabular}{|c||c|c|ccc|c|}
\hline
Dataset    & Cases & Accuracy & Mean SE    & Median SE & Std. SE    & Spearman R  \\
\hline
\hline
Training   & 59    & 0.373    & 181387.051 & 64498.932 & 384906.463 & -0.05       \\
Validation & 28    & 0.321    & 199081.864 & 73605.696 & 341473.991 & -0.011      \\
\hline
\end{tabular}
\end{table}

First, the number of evaluated cases is fairly lower when compared to the segmentation task. In fact, when training, only 164 cases had any survival information, while only $1/3$ of these cases where taken into account for the evaluation of this task. It is well-known that deep learning strategies can fail to capture enough information in the presence of a small training dataset causing issues with generalisation. In fact, we observed how the survival prediction was usually centered in the range of 200-400, while the training dataset has a range of 5-1767. Our network could not generalise using cross-validation even though we had a low number of parameters.

Second, normalising the clinical features and the subregion volumes is not a trivial task. While using the zero mean approach, the network gave negative values as output, which have no meaning for this problem. Trying to normalise between 0 and 1 did not seem to have much of an effect. Probably, because the VGG network provides a large number of features and the network (again due to the low number of training samples) was not capable to compensate.

\section{Conclusions}
\label{sec:conclusions}
In this work, we have presented a combination of CNN architectures to first segment the tumor and its subregions and then provide a survival prediction estimate using this segmentation, clinical data and image information from FLAIR, T1 and T1 post-contrast. The results in terms of segmentation, while there's room for improvement in terms of survival prediction. While using transfer learning seems to improve results over using only volume estimates, tweaking the features that are being used and how transfer learning is applied (maybe using new fully connected layers instead of the original VGG input) should improve results. We plan to further tweak our network to improve the results for the testing phase. 
 
\section*{Acknowledgments}\label{sec:Acknowledgments}
Jose Bernal and Kaisar Kushibar hold FI-DGR2017 grants from the Catalan Government with reference numbers 2017FI B00476 and 2017FI B00372 respectively. Mariano Cabezas holds a Juan de la Cierva - Incorporación grant from the Spanish Government with reference number IJCI-2016-29240. This work has been partially supported by La Fundació la Marató de TV3, by Retos de Investigació TIN2014-55710-R, TIN2015-73563-JIN and DPI2017-86696-R from the Ministerio de Ciencia y Tecnolog\'ia. The authors gratefully acknowledge the support of the NVIDIA Corporation with their donation of the TITAN-X PASCAL and TITAN-Xp GPU used in this research. The authors would like to thank the organisers of the Brats2018 challenge for providing the data.

%
%
%
\bibliographystyle{splncs04}
\bibliography{Biblio.bib}
\end{document}